\documentclass[fleqn,10pt]{SelfArx} 

\definecolor{color1}{RGB}{0,0,90} 
\definecolor{color2}{RGB}{0,20,20} 

\usepackage{hyperref} 
\hypersetup{hidelinks,colorlinks,breaklinks=true,urlcolor=color2,citecolor=color1,linkcolor=color1,bookmarksopen=false,pdftitle={Title},pdfauthor={Author},urlcolor=blue}

\usepackage{graphicx}
\usepackage{natbib}
\usepackage{amsmath}
\usepackage{multirow}
\usepackage{subfigure}
\usepackage{eucal}

\newcommand{\n}[1]{\mathrm{#1}}

\JournalInfo{Published in Journal of Magnetism and Magnetic Materials, Vol. 437, 78-85, 2017} 
\Archive{\href{http://dx.doi.org/10.1016/j.jmmm.2017.04.028}{DOI: 10.1016/j.jmmm.2017.04.028}} 

\PaperTitle{Topology optimized permanent magnet systems} 

\Authors{R. Bj\o{}rk$^*$, C. R. H. Bahl, A. R. Insinga} 
\affiliation{\textit{Department of Energy Conversion and Storage, Technical University of Denmark - DTU, Frederiksborgvej 399, DK-4000 Roskilde, Denmark}} 
\affiliation{*\textbf{Corresponding author}: rabj@dtu.dk} 

\newcommand{\keywordname}{Keywords} 

\Abstract{Topology optimization of permanent magnet systems consisting of permanent magnets, high permeability iron and air is presented. An implementation of topology optimization for magnetostatics is discussed and three examples are considered. The Halbach cylinder is topology optimized with iron and an increase of 15\% in magnetic efficiency is shown. A topology optimized structure to concentrate a homogeneous field is shown to increase the magnitude of the field by 111\%. Finally, a permanent magnet with alternating high and low field regions is topology optimized and a $\Lambda_\n{cool}$ figure of merit of 0.472 is reached, which is an increase of 100\% compared to a previous optimized design.}

\begin{document}

\flushbottom 

\maketitle 


\thispagestyle{empty} 

\section{Introduction and topology optimization}
Permanent magnets are used in a multitude of systems, ranging from motors to MRI systems \cite{Coey_2002}. In all of these applications it is important to design the permanent magnet system in such a way that it has the highest possible performance for the given application.

A permanent magnet system can be designed based on parameter variation simulations, or through use of the reciprocity theorem which for a desired linear functional gives the optimal remanence distribution \cite{Klevets_2005,Klevets_2006}. However, these approaches cannot be used to determine the optimal shape of the individual material pieces in a permanent magnet system. Although the reciprocity theorem can be used to determine the optimal border between permanent magnet and iron \cite{Insinga_2016b,Insinga_2016c}, it cannot be used to determine the optimal shape of e.g. iron pieces alone.

To determine the optimal structure of a permanent magnet system, a topology optimization approach can be utilized. Topology optimization has been used for a multitude of applications including structural mechanics, beams and trusses as well as bio-mechanical and microelectromechanical systems \cite{Bendsoe_2013}. For permanent magnet systems, topology optimization has previously been used to determine the optimal direction of the magnetization of a permanent magnet \cite{Wang_2005,Choi_2012} and for designing C-core actuators \cite{Choi_2012,Lee_2012}. Topology optimization of rotor poles in electrical motors \cite{Wang_2005,Ishikawa_2015}, in order to minimize stator slot effects \cite{Choi_2014} and the cogging torque \cite{Putek_2016}, has also been investigated. Finally, topology optimized pole pieces for MRI systems, which have high requirements on field uniformity, have been studied in both 2D \cite{Lee_2010} and 3D \cite{Tadic_2011}.

Numerically, either a finite element approach or a phase field approach is typically presented. In the latter the area to be topology optimized is discretized into pixels, each filled with a specific material \cite{Ishikawa_2015,Cheng_2011}. This allows for magnets with e.g. a specific direction of magnetization to be specified. The pixelated geometry is then refined to increase the resolution in the model.

The previous topology optimization studies of permanent magnet systems have been limited to very specific problems, typically in motor design. Topology optimization of general permanent magnet systems with varying geometrical parameters have not been considered, nor have systems where the optimal shape of bordering iron and permanent magnet pieces are considered, except for a single very specific case \cite{Lee_2012}. Here, we will consider general topology optimization of permanent magnet structures containing both permanent magnets, iron and air regions. We will consider three general permanent magnet applications and present topology optimized structures as function of various geometrical parameters for each of these. These optimized structures show a significantly improved performance compared to existing systems. First, the implementation of the topology optimization method is presented, followed by a study of a topology optimized Halbach cylinder, a topology optimized magnetic field concentrator and finally a topology optimized permanent magnet system with alternating high and low field regions.

\section{Method and implementation}
In this work, the finite element framework Comsol Multiphysics is used as the numerical implementation to perform the topology optimization simulations. The solver used is the Globally Convergent Method of Moving Asymptotes (GCMMA) solver, which is essentially a linear method with a three-level algorithm \cite{Svanberg_1987,Svanberg_2002}. This solver is ideal for problems with a large number of control variables, making it suited for topology optimization. The mesh sensitivity of the topology optimized problem will be discussed in detail subsequently.

For each topology optimization problem considered here, a global objective is defined, termed $\Theta$, that must be maximized through topology optimization. This objective must be a function of the topology of the system. The global objective of a permanent magnet system will vary depending on the application of the system. The most general optimization criteria is to obtain a permanent magnet system with the highest magnetic efficiency possible \cite{Cheng_2011}. The magnetic efficiency, or the magnetic figure of merit, is defined as \cite{Jensen_1996}
\begin{equation}\label{Eq.Mstar_definition}
M=\frac{\int_{V_\n{field}}||\mathbf{B}||^2dV}{\int_{V_\n{mag}}||\mathbf{B_\n{rem}}||^2dV},
\end{equation}
where $V_\n{field}$ is the volume of the region where the magnetic field is created and $V_\n{mag}$ is the volume of the permanent magnets. The figure of merit is the ratio of the energy stored in the field region to the maximum amount of magnetic energy available in the magnetic material, and has a maximum value of $M=0.25$ \cite{Jensen_1996}.

In a number of applications the magnetic figure of merit is not the logical choice of optimization variable. Instead the goal can e.g. be to generate as large a field as possible, regardless of the amount of permanent magnet material. Other applications such as MRI require a very uniform field, and thus here the optimization variable would be the relative standard deviation of the field.

We consider topology optimization of permanent magnet systems consisting of up to three materials: permanent magnets, high permeability iron and air. The permanent magnet material is assumed to have a linear $B-H$ relation with a permeability of $\mu=1.05$ and a fixed remanence. The high permeability iron has a non-linear $B-H$ curve as provided in the Comsol material library. The saturation magnetization is around 2 T. Previous studies on non-linear materials have shown that it is important to account for the full $B-H$ curve of the materials \cite{Lee_2012}.

In the problems considered in the following, a design region must be split into regions of two different distinct magnetic materials. In order to obtain a sharp geometrical transition between the different material types, a penalty function of a single topology control variable, $p$, is introduced that can switch between two material types. The control variable has a range between 0 and 1.

As an example, consider a design region that can either consist of permanent magnet material or high permeability iron. In this case, the relative permeability of the design region that is to be topology optimized between permanent magnet and high permeability iron is given as
\begin{equation} \label{Eq.Mur_penalty}
\mu_r = \left(1.05-\mu_{r,\n{iron}}\right)e^{-100p}+\mu_{r,\n{iron}}
\end{equation}
where $\mu_{r,\n{iron}}$ is the relative permeability of iron and the relative permeability of the permanent magnet is $\mu_{r,\n{magnet}}=1.05$. Note that $\mu_{r,\n{iron}}$ is a function of the norm of the magnetic field, $H$. Here if $p=0$ the exponential factor is 1 and the material has the relative permeability of a permanent magnet. If $p=1$, the exponential factor diminishes the first term, and the relative permeability is equal to $\mu_{r,\n{iron}}$. A somewhat similar approach was used in Refs. \cite{Choi_2014,Lee_2010}. Likewise, the remanence of the topology optimized region is given as
\begin{equation} \label{Eq.Brem_penalty}
\mathbf{B}_\n{rem} = e^{-100p}\mathbf{B}_\n{rem,\;desired}
\end{equation}
where $\mathbf{B}_\n{rem,desired}$ is the desired remanence of the permanent magnet, i.e. in the case of $p=0$ the permanent magnet has the desired remanence while for $p=1$ the remanence is essentially zero. Similarly a region can be topology optimized between high permeability iron and air by changing the factor of 1.05 in Eq. (\ref{Eq.Mur_penalty}) to 1.00, equal to the relative vacuum permeability, and setting $\mathbf{B}_\n{rem,desired}=0$ in Eq. (\ref{Eq.Brem_penalty}).

Numerical simulations have shown that the resulting topology optimized geometry depends on the initial value of $p$. This is a result of the build-in variation of the parameter in Comsol multiphysics during the optimization step. A series of simulations was conducted that determined that the optimal initial value was $p=0.003$, as this produced a topology optimized geometry with the highest performance.

We consider three different applications of topology optimization in the following. First, creating a uniform field in a cylinder bore using a Halbach cylinder is considered. Following this, the concentration of a uniform magnetic field is considered and finally generating a magnetic field with adjacent high and low field regions is considered.

\section{Halbach cylinder}
The Halbach cylinder is a cylindrical magnet system that generates a homogeneous magnetic field in the cylinder bore. This system has been used for nuclear magnetic resonance (NMR) equipment \cite{Moresi_2003,Appelt_2006}, magnetic refrigeration devices \cite{Tura_2007,Bjoerk_2010b} and medical applications \cite{Sarwar_2012}. In a continuous Halbach cylinder the components of the remanence are given in cylindrical coordinates as \cite{Halbach_1980}
\begin{eqnarray}\label{Eq.Halbach}
B_\n{rem,r} &=& B_\n{rem}\cos(p\phi)\hat{r} \nonumber\\
B_\n{rem,\phi} &=& B_\n{rem}\sin(p\phi)\hat{\phi}
\end{eqnarray}
We consider a $p=1$ cylinder that generates a homogeneous magnetic field in the cylinder bore. This Halbach cylinder has a maximum magnetic efficiency parameter of $M\approx{}0.162$, far from the theoretic maximum value of 0.25 \cite{Coey_2003,Bjoerk_2015a}. Being able to increase this value would be of great interest for all the above mentioned applications.

In general there are three ways to increase the magnetic efficiency of a given magnet design. These are using magnets with a varying norm of the remanence, altering the direction of the remanence or replacing parts of the magnetic design with a high permeability soft magnetic material. For the Halbach cylinder design, we consider a design with a constant fixed remanence. Therefore, to improve the Halbach cylinder design, it must be investigated if parts of the cylinder can be replaced with a high permeability soft magnetic material, without severely decreasing the produced field. The goal of a topology optimized Halbach cylinder is to improve the magnetic efficiency of the design such that a higher magnetic field could be generated using less amount of magnet.

\begin{figure*}[!p]
\centering
\subfigure[]{\includegraphics[width=1\columnwidth]{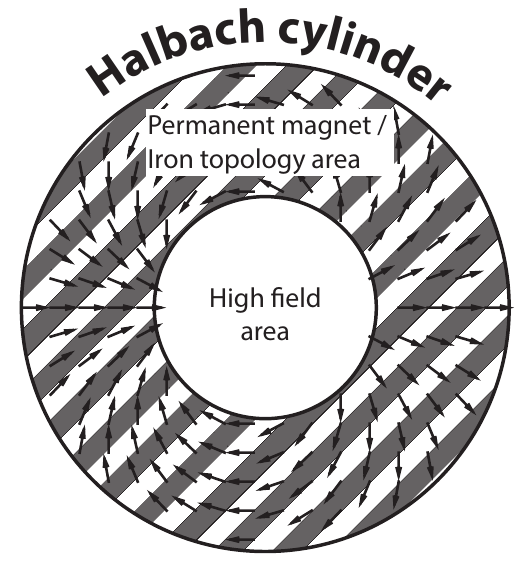}}
\subfigure[]{\includegraphics[width=1\columnwidth]{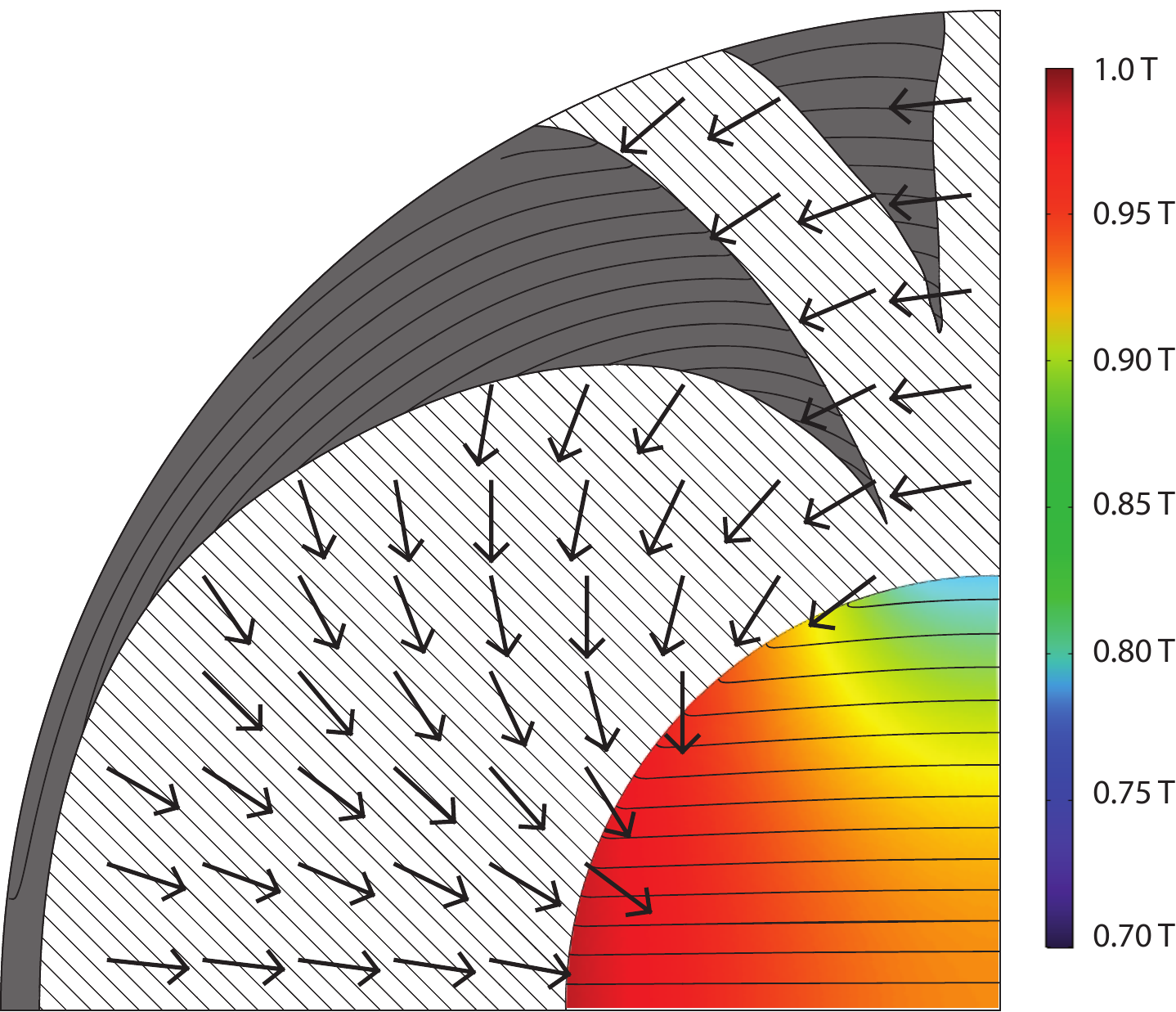}}
\caption{a) The Halbach geometry to be topology optimized. The hatched grey area indicates the area to be topology optimized. The arrows indicate the fixed direction of remanence, if the material in question is a permanent magnet. b) A quarter of the topology optimized system for $R_\n{o}/R_\n{i}=2.3$.}\label{Fig.Halbach_ill}
\end{figure*}

The whole cylindrical Halbach magnet is considered as the topology optimization area, as shown in Fig. \ref{Fig.Halbach_ill}. The material in a point on the cylinder is either permanent magnet with a remanence given by Eq. (\ref{Eq.Halbach}) or iron with soft magnetic properties as stated previously. The objective function that is maximized is
\begin{equation}\label{Eq.Theta_Halbach}
\Theta = \frac{\langle{}B\rangle{}}{A_\n{mag}}
\end{equation}
where $\langle{}B\rangle{}$ is the average magnetic flux density in the cylinder bore and $A_\n{mag}$ is the area of the permanent magnet, i.e. the area of the cylinder that is not iron.

\begin{figure*}[!p]
  \centering
  \includegraphics[width=2\columnwidth]{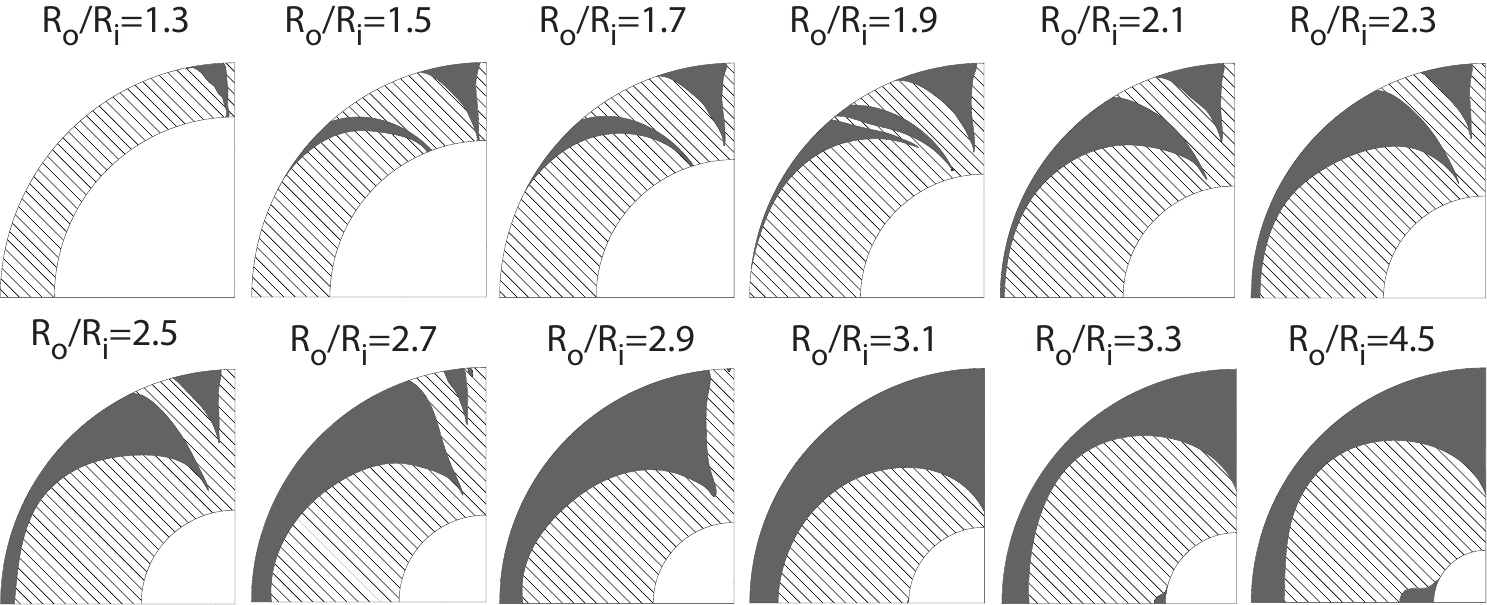}
  \caption{The topology optimized geometries as function of the ratio between the outer and the inner radius. The grey areas are iron and the hatched areas are permanent magnet with a remanence given by Eq. \ref{Eq.Halbach}. Only a quarter of the designs are shown, as the designs are mirror symmetric.}
    \label{Fig.Illus_All_arrows_simplified}
\end{figure*}

For the Halbach cylinder, the norm of the remanence is constant as is the area of the high field region, i.e. the cylinder bore. This means that the figure of merit, $M$, reduces to a proportionality between the integral of the magnetic flux density squared and the volume of the magnet squared. However, in practical applications it is usually required that the Halbach cylinder generate as strong a field as possible, using the least amount of magnetic material. This corresponds to the optimization parameter $\Theta$ defined in Eq. \ref{Eq.Theta_Halbach}. If the generated field is completely homogeneous, maximizing $\Theta$ also maximizes $M$.

The geometry resulting from the topology optimization process depend on the ratio between the outer and the inner radius of the Halbach cylinder, $R_\n{o}$ and $R_\n{i}$ respectively. The computed topology optimized geometries are shown in Fig. \ref{Fig.Illus_All_arrows_simplified}. As can be seen from the figure, the fractional area of iron increases as function of the ratio of the outer and inner radius. The topology optimization produces iron regions with features with very sharp ends. These are similar to structures seen by Ref. \cite{Choi_2014}, although in that case it was for borders between iron and air, and not iron and permanent magnet as is the case here.

\begin{figure*}[!t]
\centering
\subfigure[]{\includegraphics[width=1\columnwidth]{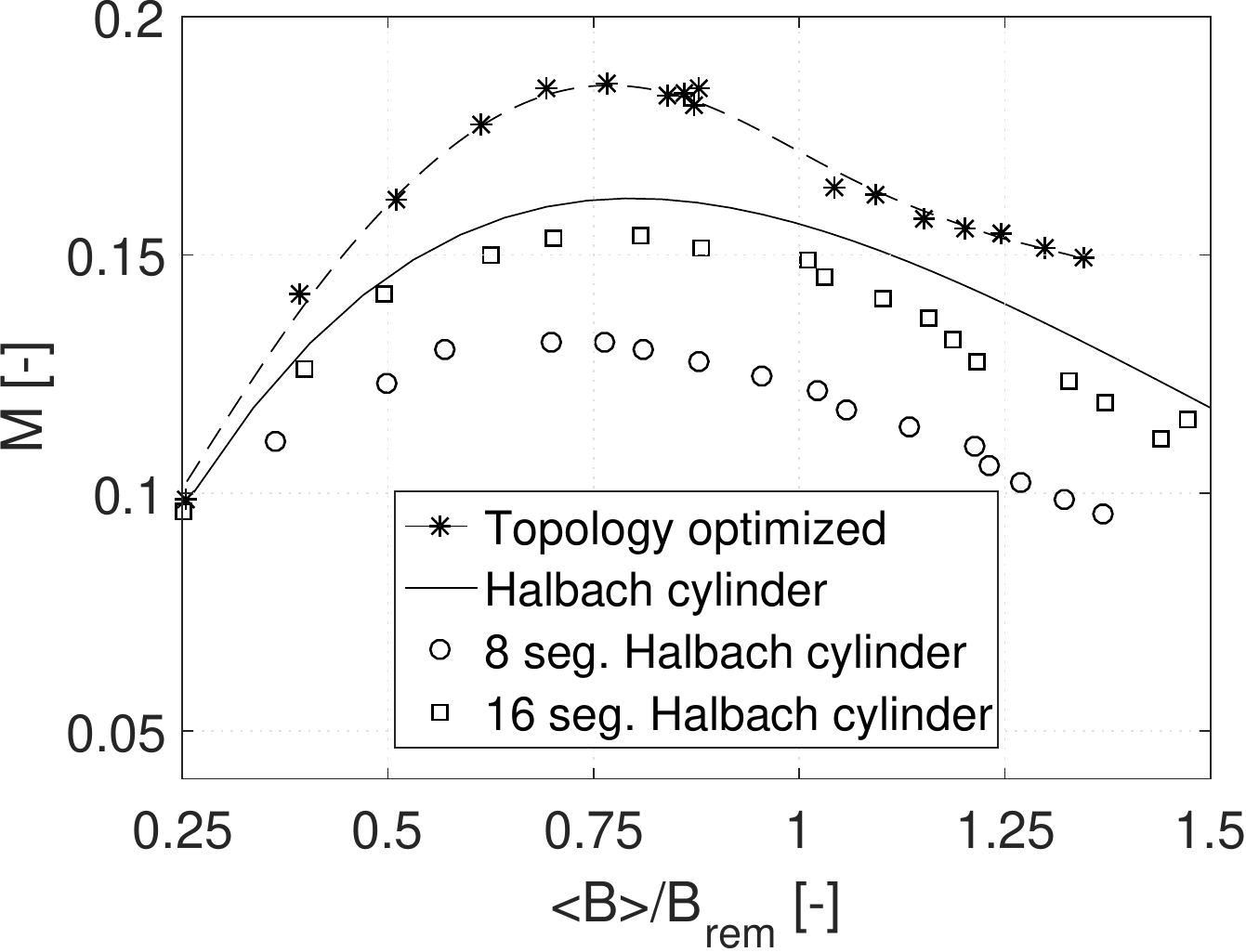}}
\subfigure[]{\includegraphics[width=1\columnwidth]{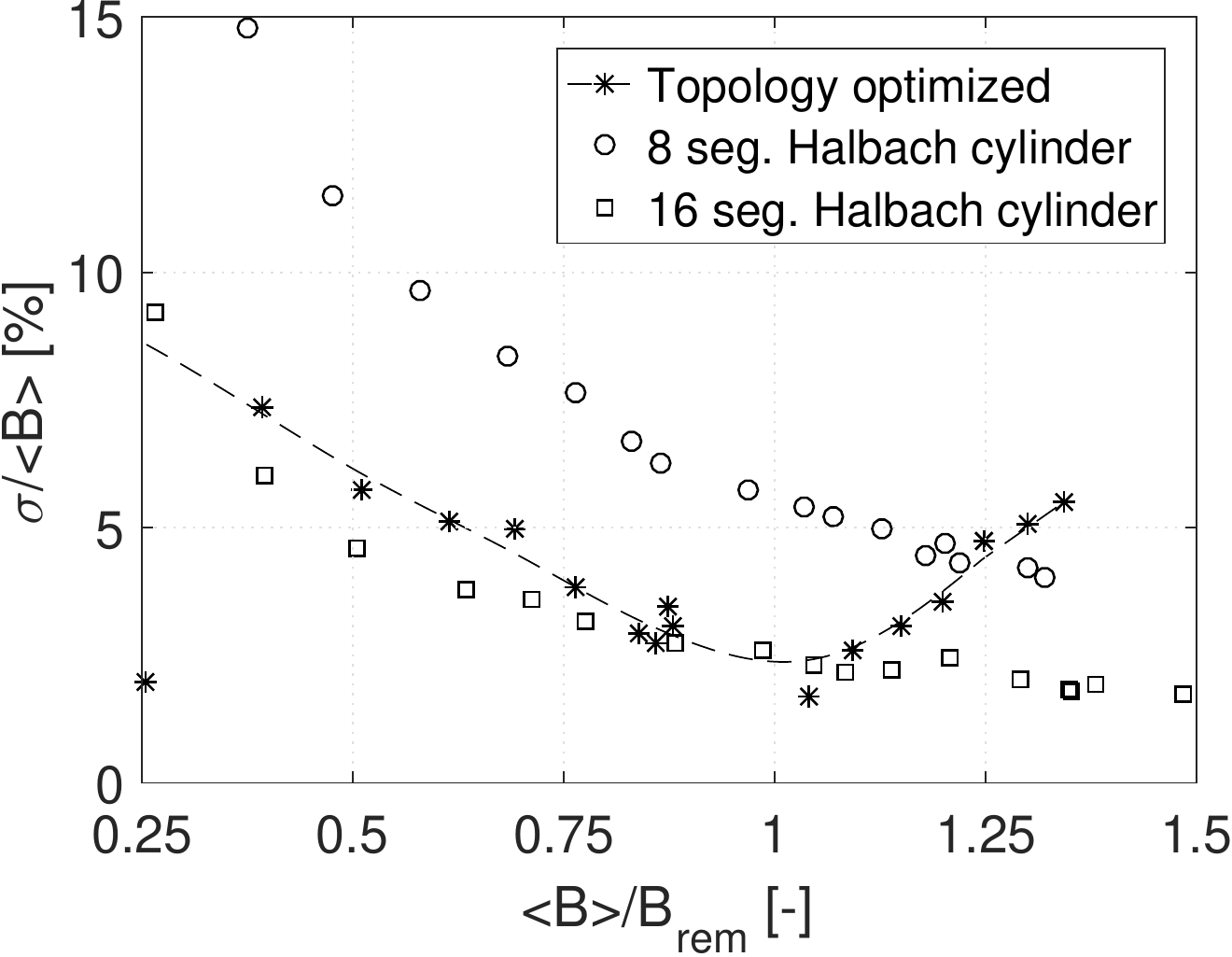}}
\caption{a) The figure of merit, $M$, and b) the relative standard deviation of the field generated in the cylinder bore, both as function of the average field generated in the bore normalized by the remanence. The lines are guides to the eye.}\label{Fig.Halbach_topop}
\end{figure*}

The figure of merit, $M$, for the topology optimized designs is shown in Fig. \ref{Fig.Halbach_topop}a as function of $\langle{}B\rangle{}/B_\n{rem}$. For the Halbach cylinder, the figure of merit can explicitly be calculated \cite{Jensen_1996,Bjoerk_2015a}. This expression is also shown in the figure. As can be seen from the figure, the topology optimized design is superior to the Halbach cylinder for all values of the field generated. The generated field increases monotonically with an increase in $R_\n{o}/R_\n{i}$. However, the stronger field generated comes at the expense of the homogeneity of the field in the bore. The relative standard deviation of the field in the bore as a function of the average norm of the field is shown in Fig. \ref{Fig.Halbach_topop}b. The relative standard deviation is given as $\frac{\sigma}{\langle{}B\rangle{}}=\frac{\sqrt{\langle{}(B-\langle{}B\rangle{})^2\rangle{}}}{\langle{}B\rangle{}}$. The field is quite homogeneous with a relative standard deviation below 8\% for all fields considered. The large change of the standard deviation at the data point at $\langle{}B\rangle{}/B_\n{rem}=0.25$ is due to a change in the topology of the system from one to two iron regions, as can also be seen in Fig. \ref{Fig.Illus_All_arrows_simplified} from $R_\n{o}/R_\n{i}=1.3$ to $1.5$. Note that for a Halbach cylinder, the field is completely homogeneous and the standard deviation is zero. The standard deviation of a segmented 8 or 16 piece Halbach cylinder is also shown, and the topology optimized structure is seen to have a relative standard deviation very similar to the 16 segmented Halbach.

As an example, at the optimal Halbach efficiency of $\langle{}B\rangle{}/B_\n{rem}=0.796$, the figure of merit is $M=0.186$ for the topology optimized design, an improvement of 15\% compared to the Halbach cylinder. This improvement comes at the expense of a small increase in relative standard deviation of the field in the bore to $\frac{\sigma}{\langle{}B\rangle{}}=3.8\%$. This system has $R_\n{o}/R_\n{i}=2.3$ and is illustrated in Fig. \ref{Fig.Halbach_ill}b.

\section{Concentrating a homogeneous field}
We now consider a homogeneous magnetic field across a region of space and wish to see if a structure that concentrates the magnetic field in a given area can be designed using topology optimization. This could be a device similar to the meta-material flux enhancer presented in Ref. \cite{Navau_2012}, except using purely ferromagnetic material. In this system the choice between materials is iron and air, contrary to the choice between permanent magnet and iron in the Halbach cylinder case considered previously.

\begin{figure*}[!t]
\centering
\subfigure[]{\includegraphics[width=1\columnwidth]{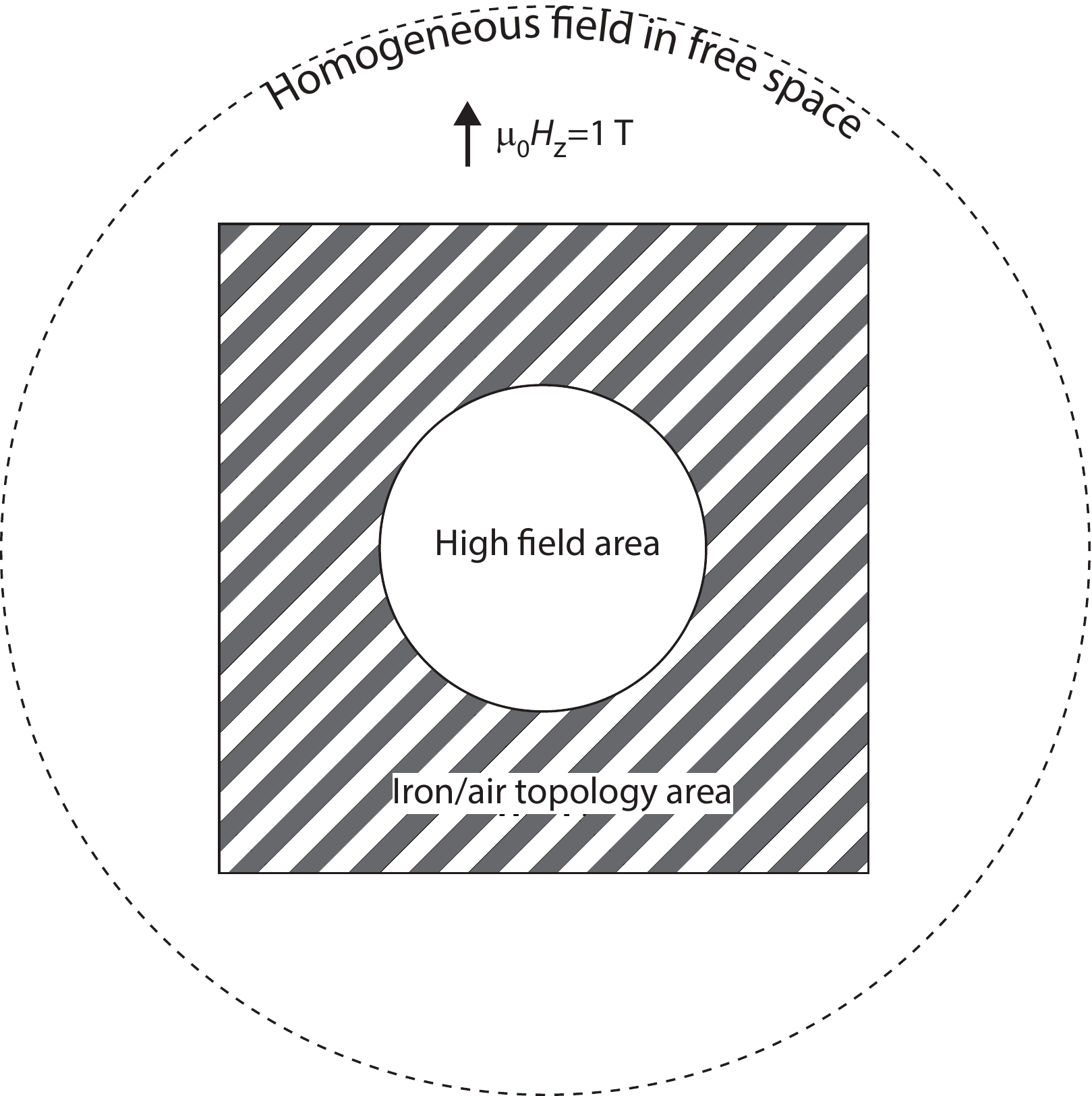}}
\subfigure[]{\includegraphics[width=1\columnwidth]{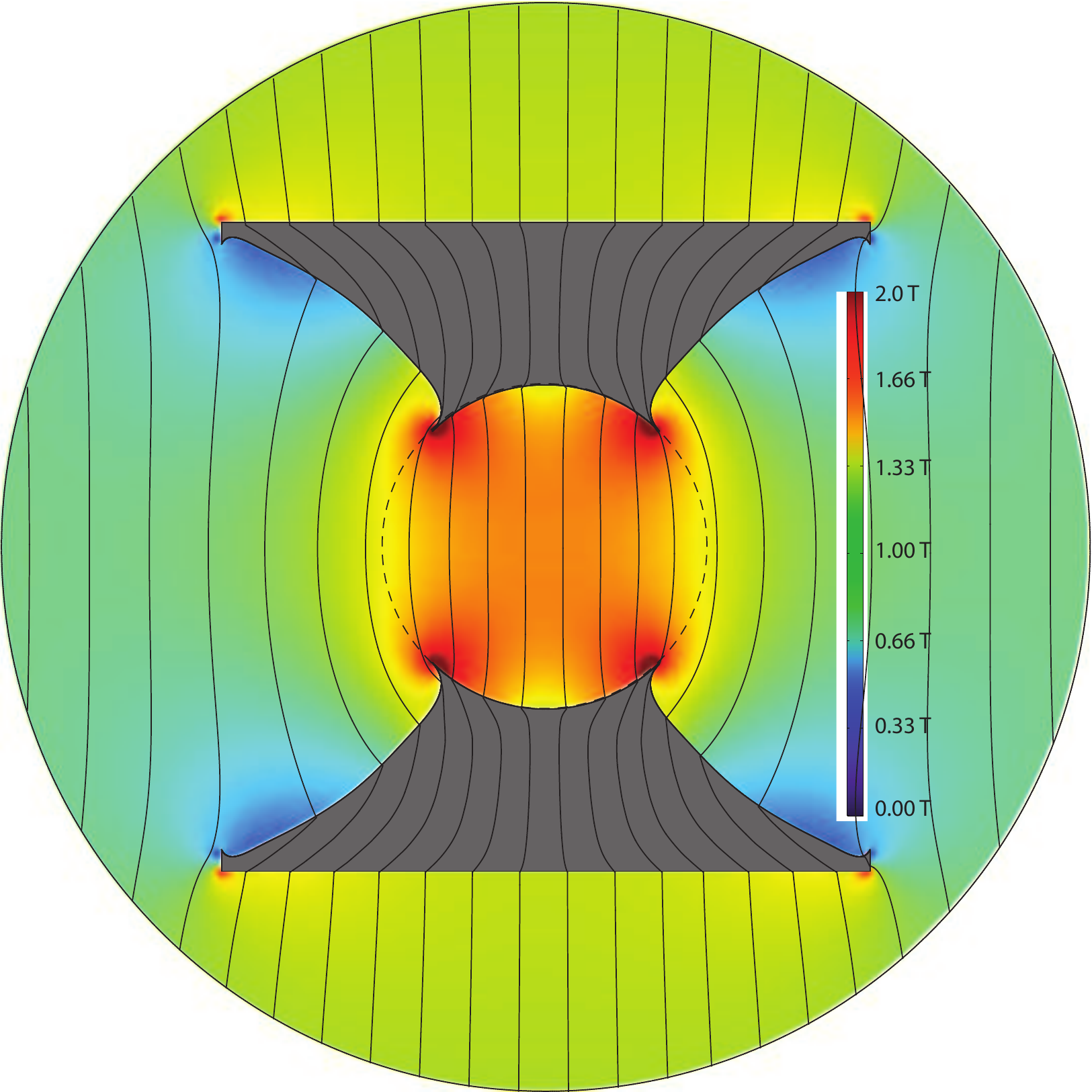}}
\caption{a) The geometry of the system considered. The hatched grey area indicate the area to be topology optimized. b) The topology optimized structure determined.}\label{Fig.Homo_ill}
\end{figure*}

We consider a two-dimensional geometry as shown in Fig. \ref{Fig.Homo_ill}a. The area to be topology optimized is shaped as a square, surrounding a circular high field region. The diameter of the circular high field region is half the side length of the surround square. The area to be topology optimized can in a given point have a permeability in the range of $\mu_r=1$ to 4000, i.e. from free space to that of unsaturated iron. Following the computation, the computed geometry is verified with the actual nonlinear $B-H$ of iron. The topology optimization criteria, $\Theta$, is a maximization of the average field throughout the bore, i.e. without regard for field uniformity,
\begin{equation}
\Theta = \langle{}B\rangle{}
\end{equation}

We can estimate the amount that the magnetic field can at most be concentrated by following the approach of Ref. \cite{Navau_2012}. Although this approach has only been shown to be valid for circular geometries, it will never the less provide an estimate for the geometry considered here. In this framework the increase in field is proportional to the difference in cross-sectional length of the geometry perpendicular to the magnetic field. Here this is the side length of the square topology area divided by the diameter of high field region, i.e. a factor of two.

The computed topology optimized structure is shown in Fig. \ref{Fig.Homo_ill}b using the actual nonlinear $B-H$ curve. In this structure the field is enhanced from a surrounding homogeneous 1 T magnetic field to an average field of $\langle{}B\rangle{} = 1.46$ T in the high field area, an increase of 46\%. The field generated is not uniform, but this was also not a requirement of the optimization algorithm. The increase in field of 46\% is less than the maximum possible 100\% increase. However, as only material with $\mu_\n{r} \geq 1$ is used, it is not possible to completely shield the system from flux leakage.

\begin{figure}[!tp]
  \centering
  \includegraphics[width=1\columnwidth]{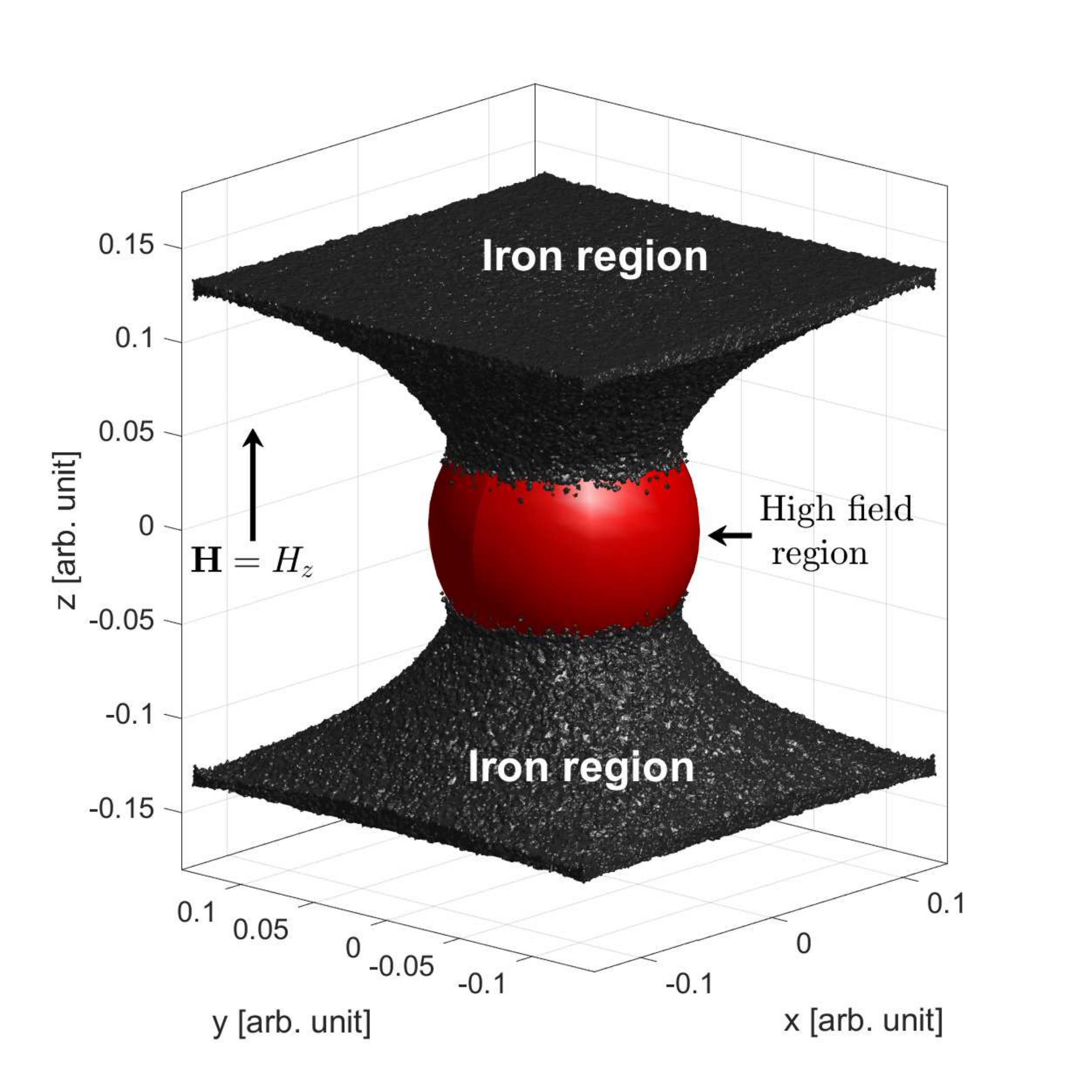}
  \caption{The geometry of the 3D topology optimized structure. The iron regions as well as the high field region are indicated. The applied field is $\mu_{0}H_\n{z}=1$ T.}
    \label{Fig.Homo_3D_top}
\end{figure}

This structure was also optimized in 3D, where the high field region was a sphere and the topology optimization region surround the sphere was shaped as a cube. The resulting structure in this case was a structure similar to that in Fig. \ref{Fig.Homo_ill}b, except rotated around the sphere, as shown in Fig. \ref{Fig.Homo_3D_top}. Here the averaged field was increased to $\langle{}B\rangle{}=2.11$ T, an increase of 111\% compared to the surrounding field.

Here the same approach as above leads to a maximum concentrating factor of $8/\pi=2.54$, i.e. the side length of the iron region divided by the circumference of the high field region. Again, a smaller increase than the theoretical limit is seen, due to the same reasons as argued above.

\subsection{Pole pieces}
\begin{figure}[!t]
  \centering
  \includegraphics[width=\columnwidth]{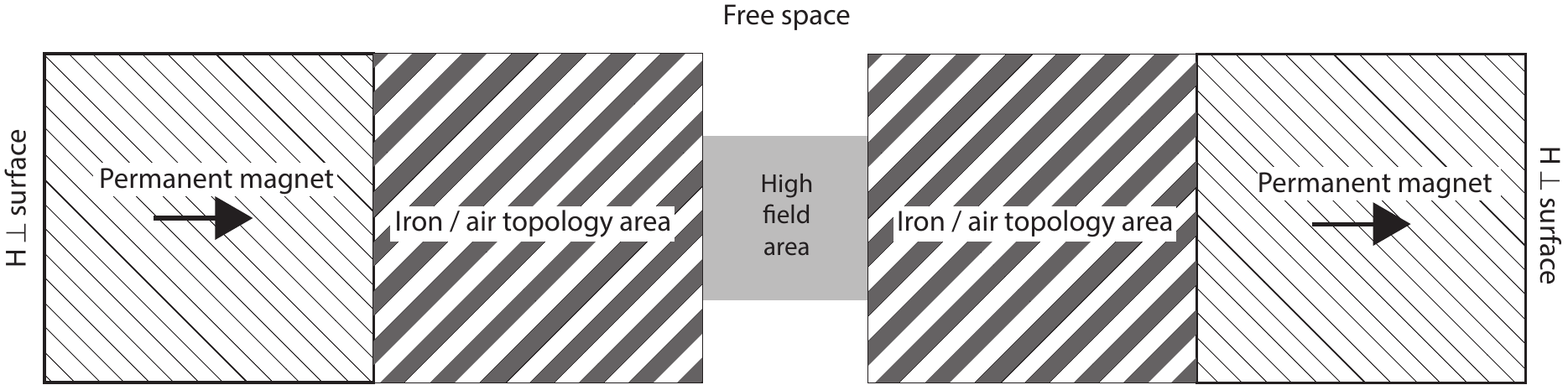}
  \caption{The pole piece geometry simulated. The hatched grey area indicate the areas to be topology optimized. The two permanent magnets are considered in contact through an iron yoke (not shown).}
    \label{Fig.Pole_piece_ill}
\end{figure}

We now extend the above analysis to a system that includes permanent magnets as the flux sources. We consider a geometry as shown in Fig. \ref{Fig.Pole_piece_ill}, i.e. a system with pole pieces to focus and enhance a magnetic field. The geometry consists of two square permanent magnets, two regions that are to be topology optimized between air and iron and a smaller square area where the magnetic field is to be maximized. The side length of the high field region is half of the side length of the permanent magnets. The flux lines are imagined closed between the two permanent magnets through an iron circuit. Numerically, this is accomplished through periodic boundary conditions.

We consider a permanent magnet with a remanence of 1 T. We wish to determine the topological shape of iron that can concentrate the magnetic field into the high field area. Contrary to the case above, both a strong and homogeneous field is desired. Therefore, the optimization function to be maximized is given as 
\begin{equation}
\Theta = \frac{{\langle{}B\rangle{}}^\delta}{\sqrt{\langle{}(B-\langle{}B\rangle{})^2\rangle{}}}
\end{equation}
In this expression the nominator is the average norm of the field, while the denominator is the standard deviation of the field. The factor $\delta$ is used to prioritize a strong but inhomogeneous field by increasing the absolute value of the nominator.

In order to evaluate the effectiveness of the topology optimization algorithm, we compare the generated topologies with a standard trapezoidal-shaped pole piece, i.e. a geometry where the area to be topology optimized is replaced by a trapezoid of iron with varying side angle. Interestingly, the magnetic field produced in the air gap by the trapezoid geometry can also be computed using a simple magnetic circuit approach for this geometry, including flux leakage in the model \cite{Leupold_1996}. There is a difference of less than 10\% between finite element calculations and the magnetic circuit model for the geometry.

\begin{figure*}[!t]
\centering
\subfigure[]{\includegraphics[width=1\columnwidth]{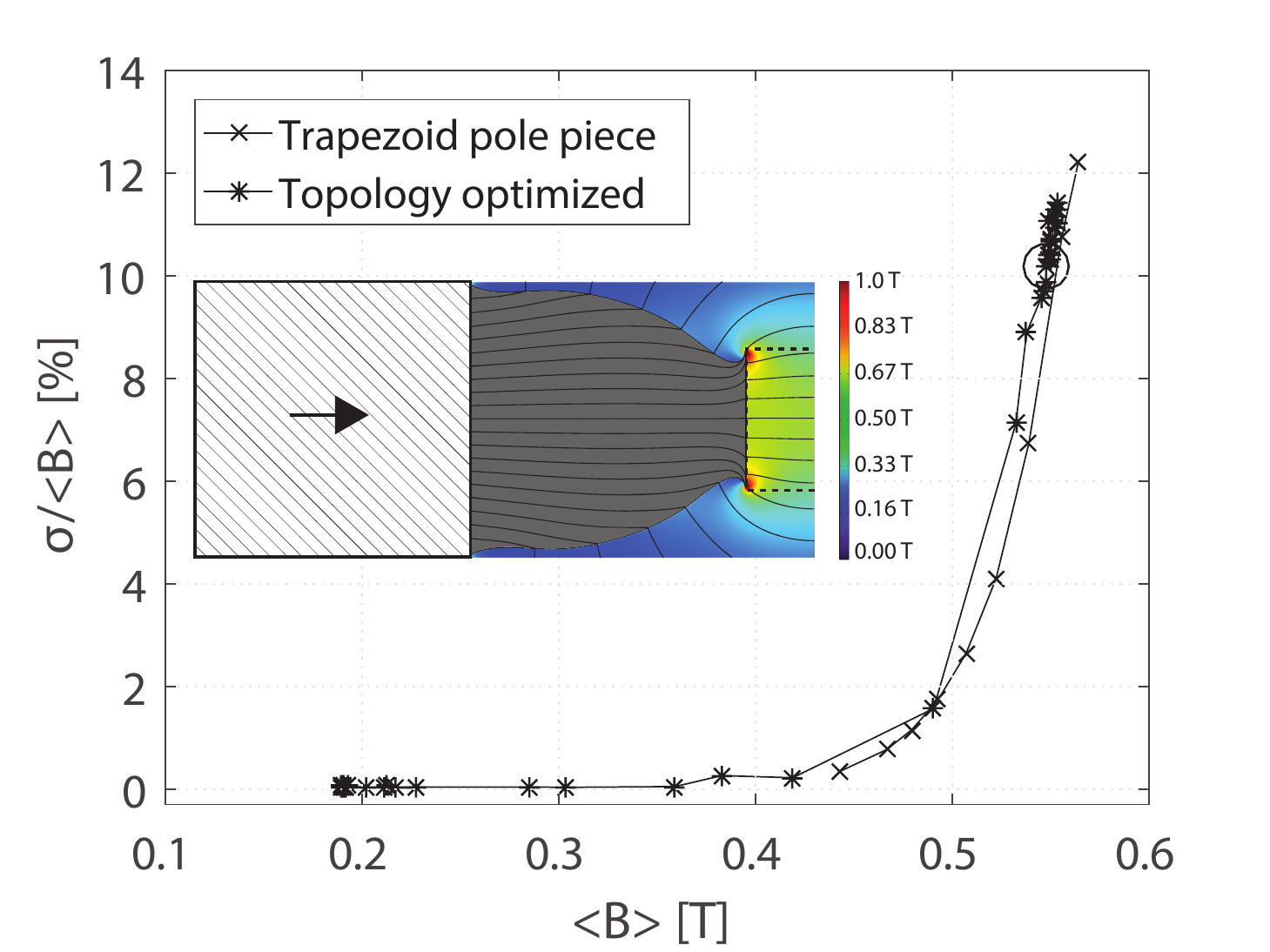}}
\subfigure[]{\includegraphics[width=1\columnwidth]{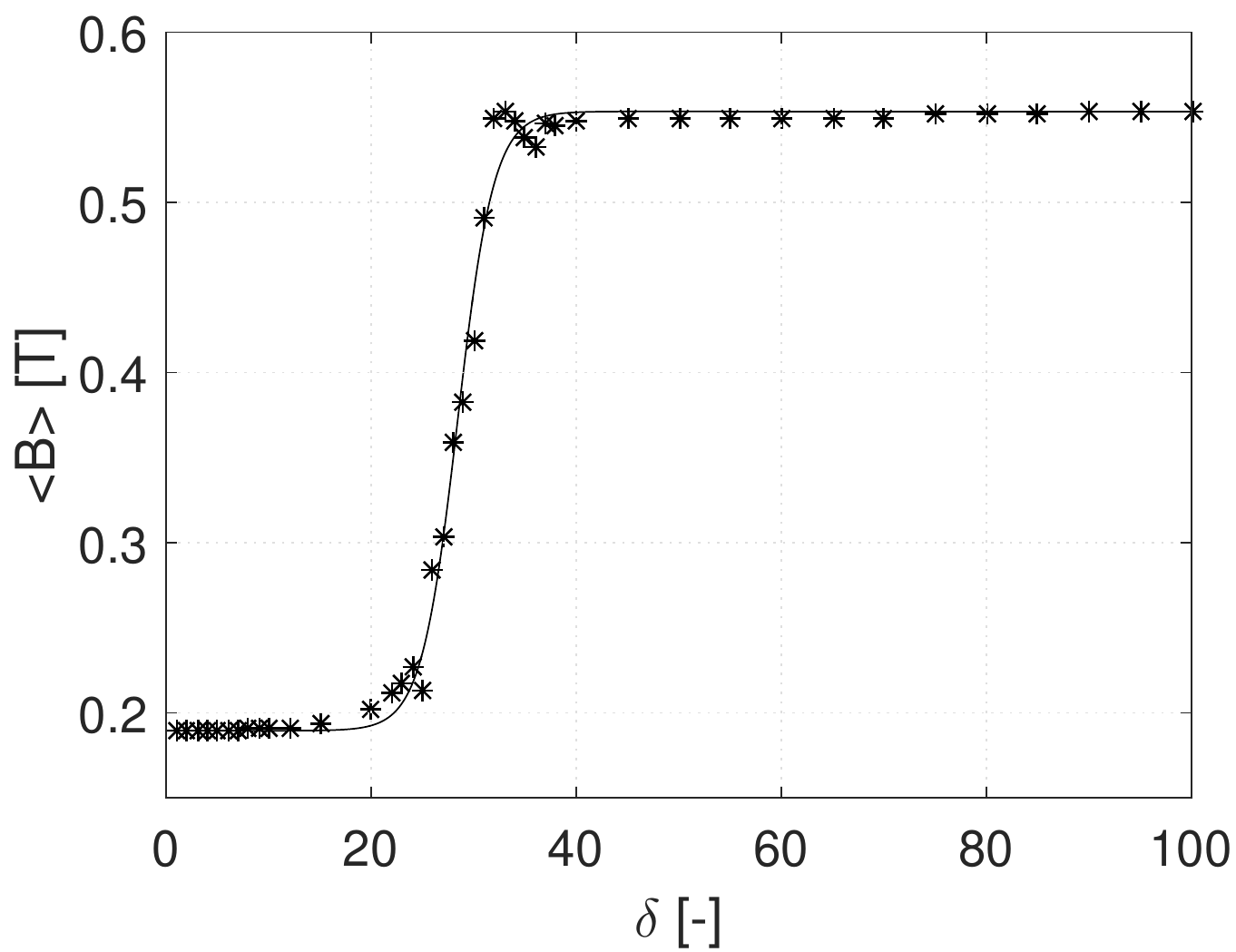}}
\caption{a) The homogeneity of the field in the high field area as function of the average norm of the field in the same area. Both the results from topology optimized geometries as well as for a trapezoidal shape pole piece are shown. The inset shows the geometry for $\delta=40$, which is the data point marked with a circle. b) The average norm of the magnetic field as function of the $\delta$ parameter. The line is a guide to the eye.}\label{Fig.Pole_piece_max_and_homo}
\end{figure*}

The relative standard deviation of the field in the high field area as function of $\langle{}B\rangle{}$ is shown in Fig. \ref{Fig.Pole_piece_max_and_homo}a for both the topology optimized geometries as well as for the trapezoidal geometries. The parameter $\delta$ is varied to control the strength of the generated field. As can be seen from the figure, the topology optimization algorithm is not able to determine a geometry that is better than the simple trapezoidal shape for the considered geometry. The dependence of the topology optimization on the $\delta$ parameter can be seen in Fig. \ref{Fig.Pole_piece_max_and_homo}b. Here it is clearly seen that as $\delta$ is increased, a higher field in the high field region is prioritized, at the expense of the homogeneity of the field.

\section{Magnetic refrigeration}
We now consider permanent magnet structures, for which the low field regions are also of importance. This is relevant for magnetic refrigeration, where the permanent magnet system must provide adjacent regions of high and low field, between which a magnetocaloric material can be moved. Realizing this in practice without substantial flux leakage is difficult, although a number of magnet designs have realized a large difference between high and low field regions \cite{Bjoerk_2010b}.

\begin{figure*}[!t]
\centering
\centering
  \includegraphics[width=2\columnwidth]{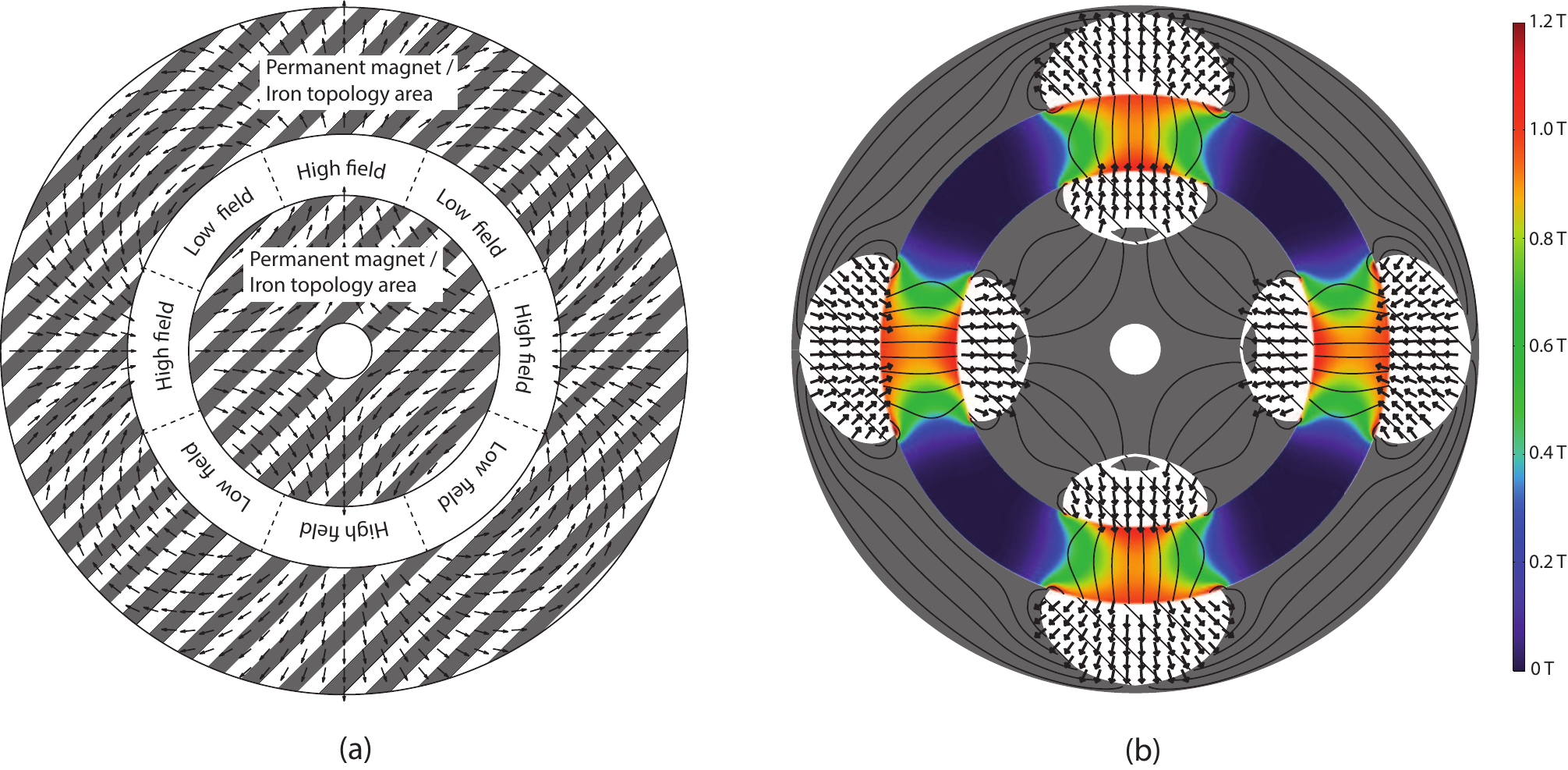}
\caption{a) The geometry of the system considered. b) The topology optimized structure determined, with $B_\n{rem}=1.4$ T.}\label{Fig.Mag_Refri_ill}
\end{figure*}

To more easily compare a topology optimized geometry with previous results, we consider a permanent magnet system geometry previously optimized using an alternative approach \cite{Bjoerk_2010f,Bjoerk_2011c}. This geometry consists of two concentric Halbach cylinders, that generate four high field and four low field regions in the space between the cylinders. The radii of the system are identical to those given in Ref. \cite{Bjoerk_2010f}, namely an inner and outer radius of the inner magnet of 10 mm and 70 mm respectively, and corresponding radii of the outer magnet of 100 mm and 135 mm, respectively.  The remanence of the outer magnet is given by Eq. \ref{Eq.Halbach} with $p=2$, while the inner magnet has $p=-2$. The geometry is illustrated in Fig. \ref{Fig.Mag_Refri_ill}a. The topology optimization routine will here distinguish between permanent magnet material and iron.

The optimization criteria must be designed to favor either a large difference in magnetic field between the high and low field regions or a small amount of permanent magnet used to create this field difference. The expression to be maximized is
\begin{equation}\label{Eq.Opt_mag_refri}
\Theta = \delta^{\langle{}B_\n{high}^{2/3}\rangle{}-\langle{}B_\n{low}^{2/3}\rangle{}}\frac{A_\n{field}}{A_\n{mag}}
\end{equation}
Here $\langle{}B_\n{high}^{2/3}\rangle{}$ is the average of the magnetic field to the power of $2/3$ in the high field region, and similarly for the low field region, and $A_\n{field}$ and $A_\n{mag}$ are the areas of the high field volume and the area of permanent magnet material, respectively. Note that as the geometry is fixed the area of the high field region is $A_\n{field}=\pi/2\left((100\;\n{mm})^2-(70\;\n{mm})^2\right)$. The power of $2/3$ is used because the magnetocaloric effect approximately scales with this power around Curie temperature for most relevant magnetocaloric materials \cite{Bjoerk_2010d,Smith_2012}. The parameter $\delta$ can be increased to increase the difference in field strength produced by the magnet, albeit with a higher amount of permanent magnet material.

\begin{figure*}[!t]
\centering
\subfigure[]{\includegraphics[width=1\columnwidth]{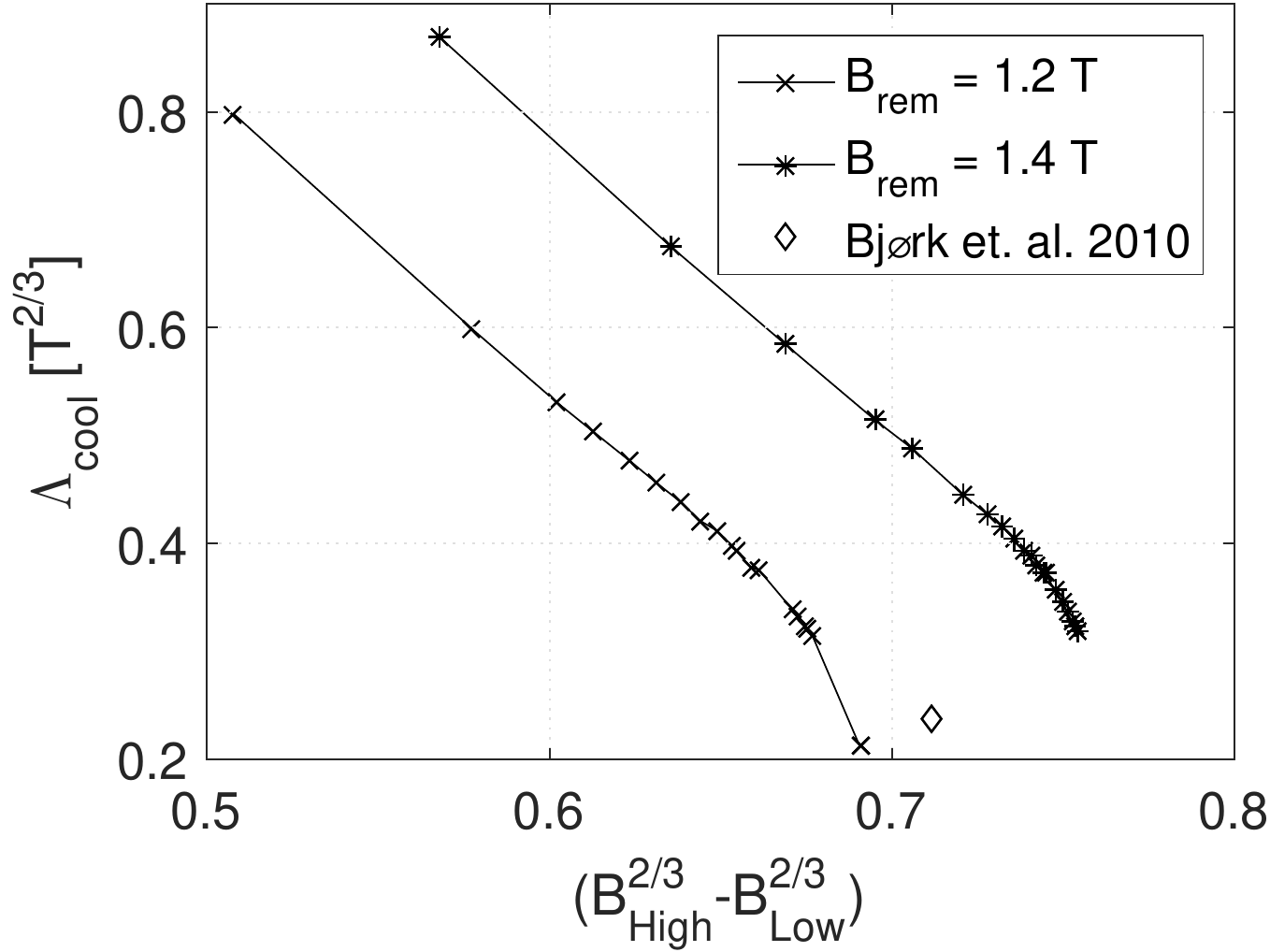}}
\subfigure[]{\includegraphics[width=1\columnwidth]{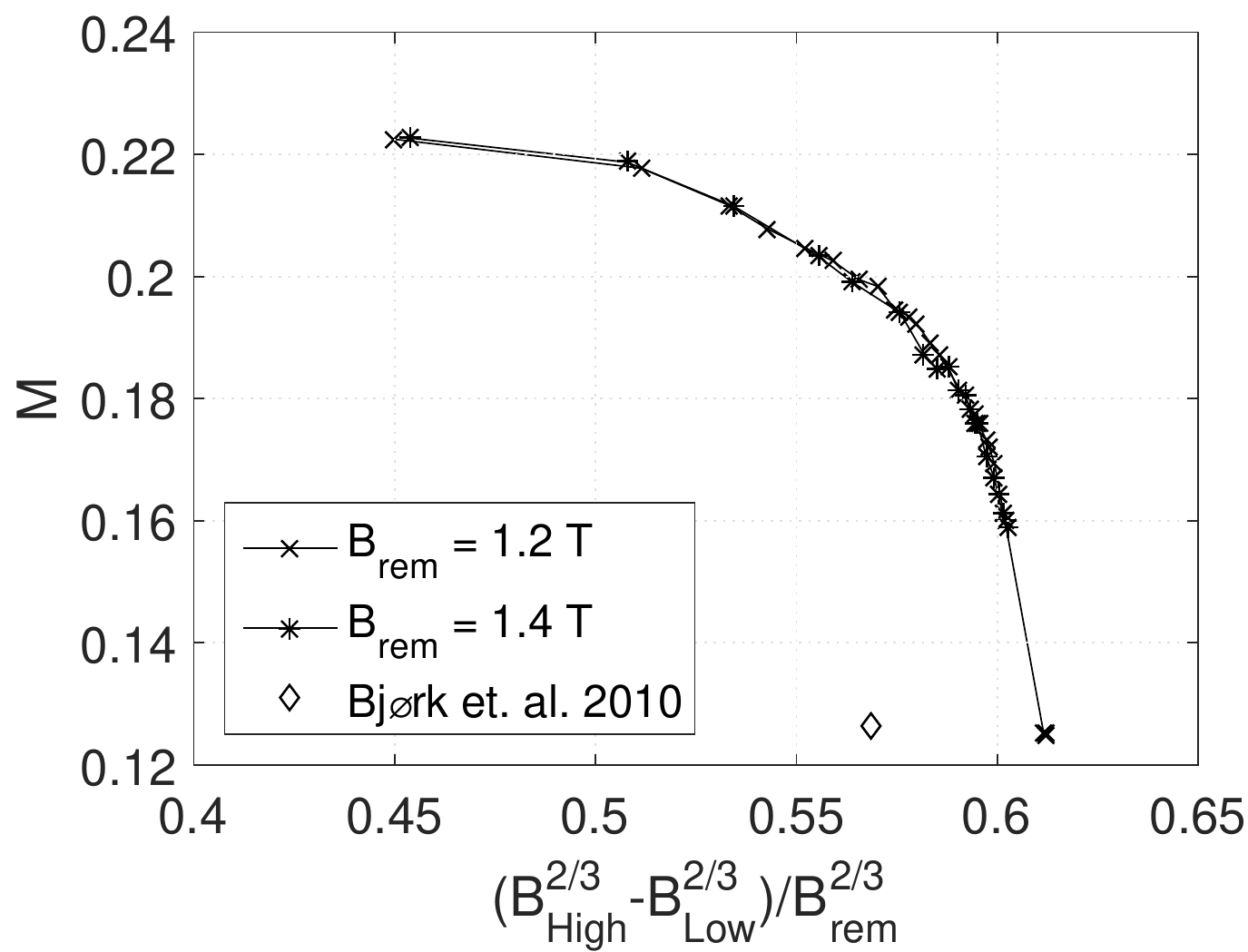}}
\caption{The increase in a) $\Lambda_\n{cool}$ and b) the figure of merit, $M$, as function of the difference in field to the power of 2/3.}\label{Fig.Mag_Refri}
\end{figure*}

The geometry is topology optimized for $\delta{}=1-50$. An example of one of the topology optimized structures is shown in Fig. \ref{Fig.Mag_Refri_ill}b. The efficiency of the resulting designs can be determined by calculating the $\Lambda_\n{cool}$ parameter \cite{Bjoerk_2008}, which is proportional to Eq. (\ref{Eq.Opt_mag_refri}) except without $\delta$ in the first factor. The $\Lambda_\n{cool}$ and the $M$ figure of merit value for the topology optimized designs are shown in Fig. \ref{Fig.Mag_Refri} as function of the difference in field to the power of $2/3$. Two different remanence values of 1.2 T and 1.4 T were used, as $\Lambda_\n{cool}$ does not account for the influence of this parameter. The topology optimized structures are compared with the structure optimized using conventional optimization \cite{Bjoerk_2010f} and with $B_\n{rem}=1.4$ T. As can be seen from the figure, the topology optimized structures are significantly better than the conventionally optimized structure. With regards to the magnetic efficiency, $M$, the structures with different but constant remanence perform identically, as expected.

It should be noted that the low difference in flux density is a result of the dimensions of the geometry. Had larger magnets been utilized, the difference in flux density would have increased substantially.

\section{Discussion}
The objective function as well as the resulting improvement for each of the above cases is shown in Table \ref{Table.Cases}. As can clearly be seen, topology optimization can be used to design permanent magnet systems with significantly increased performance, compared to existing designs.
\begin{table*}\begin{center}
\caption{\label{Table.Cases}The value for the Halbach cylinder is taken at $R_\n{o}/R_\n{i}=2.3$, while the value for magnetic refrigeration is taken at $\langle{}B_\n{high}^{2/3}\rangle{}-\langle{}B_\n{low}^{2/3}\rangle{}=0.71$ T$^{2/3}$.}
\begin{tabular}{l|c|c|c|c}
   Type                   & Objective          & Optimal   & Improvement   \\
                          & function, $\Theta$ & value     & (reference)  \\ \hline
   Halbach cylinder       & $\frac{\langle{}B\rangle{}}{V_\n{mag}}$  &      $M=0.186$    & 15\% \cite{Coey_2003}     \\
   Homogeneous field      & $\langle{}B\rangle{}$  &  $\langle{}B\rangle{}=2.11$ T                  & 111\%     \\
   Pole pieces            & $\frac{{\langle{}B\rangle{}}^\delta}{\sqrt{\langle{}(B-\langle{}B\rangle{})\rangle{}^2}}$     &   Identical to trapeziodal pole piece  & None     \\
   Magnetic refrigeration & $\delta^{\langle{}B_\n{high}^{2/3}\rangle{}-\langle{}B_\n{low}^{2/3}\rangle{}}\frac{A_\n{field}}{A_\n{mag}}$     &  $\Lambda_\n{cool}=0.472$ T$^{2/3}$ & 100\% \cite{Bjoerk_2010f}
\end{tabular}\end{center}
\end{table*}

In the above implementation of topology optimization for permanent magnets, we have considered permanent magnets with a specified distribution of remanence. It is possible also to consider topology optimized problems where the remanence can freely vary in orientation throughout the design region by introducing a control variable describing e.g. the angle of the remanence with respect the chosen coordinate system. However, this is computationally very intensive and therefore such problems have not been considered above.

For the calculations above self-demagnetization has not been considered, but as long as the assemblies remain small \cite{Bjoerk_2015a} or magnets with high coercivity are used \cite{Insinga_2016a}, this is not necessary. Furthermore, the manufacturability or the designs has not been considered. Numerical methods exist to optimally segment permanent magnet structures \cite{Insinga_2016b}, which can be applied to segment and reduce the complexity of the topology produced systems for different applications \cite{Insinga_2016c}.

\begin{figure}[!tp]
  \centering
  \includegraphics[width=1\columnwidth]{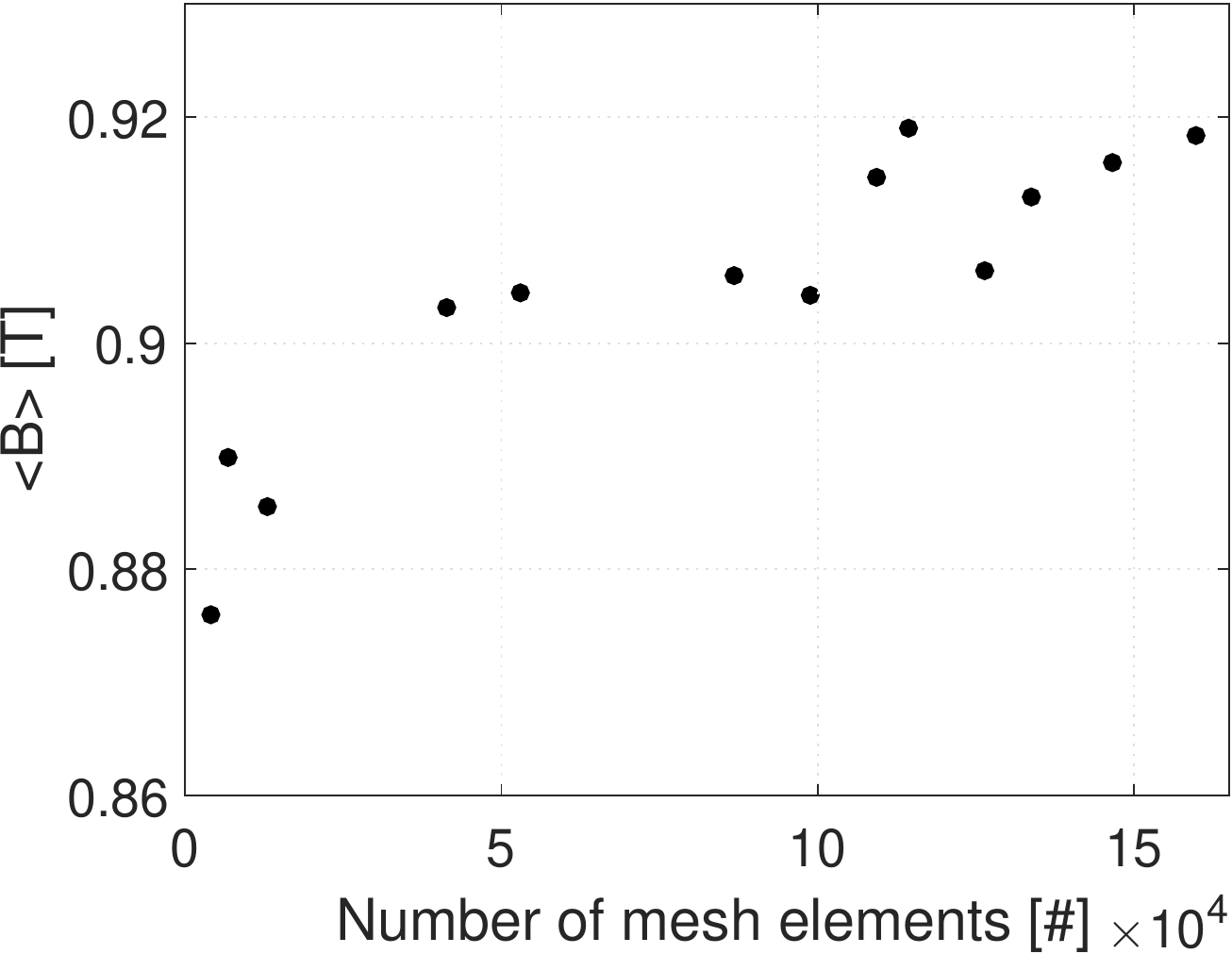}
  \caption{The average value of the flux density in a topology optimized Halbach cylinder with a ratio of the inner and outer radius of $R_\n{o}/R_\n{i}=2.3$ as function of the number of triangular elements in the mesh.}
    \label{Fig.Mesh_analysis}
\end{figure}

The mesh sensitivity of the topology optimized problems considered were investigated using the Halbach cylinder geometry defined above. The average field in the cylinder bore as function of the number of mesh elements is shown in Fig. \ref{Fig.Mesh_analysis}. As can be seen from the figure, the solution clearly converge to a fixed value. The choice for number of mesh elements used for all simulations throughout this article corresponds to the highest number of mesh elements shown in Fig. \ref{Fig.Mesh_analysis}.

\section{Conclusion}
In conclusion, we have presented topology optimization of permanent magnet structures consisting of permanent magnets, high permeability iron and air. Three examples were considered. First, the Halbach cylinder was topology optimized by inserting iron, and an increase of 15\% in magnetic efficiency was obtained with only an increase of 3.8 pp. in field inhomogeneity - a value similar to the inhomogeneity in a 16 segmented Halbach cylinder. Following this a topology optimized structure to concentrate a homogeneous field was computed, and was shown to increase the magnitude of the field by 111\%. Finally, a permanent magnet with alternating high and low field regions was considered. Here a $\Lambda_\n{cool}$ figure of merit of 0.472 was obtained, which is an increase of 100\% compared to a previous optimized design.

\section*{Acknowledgements}
This work was financed by the ENOVHEAT project, which is funded by Innovation Fund Denmark (contract no 12-132673).

\bibliographystyle{elsarticle-harv}

\end{document}